\documentclass[11pt,a4paper,twocolumn,final,numbers,sort,sort&compress]{revtex4}

\usepackage{amsfonts,amssymb,amsmath}     
\usepackage{latexsym}                     
\usepackage{mathrsfs,bm}                  

\usepackage{graphicx,pst-all,epsfig}      

\usepackage[utf8]{inputenc}               

\usepackage{float,afterpage}              
\usepackage{longtable}                    
\usepackage{hyperref}                     

\usepackage{fancyhdr}                     
\usepackage{lastpage}                     
\usepackage{setspace}                     

\begin{document}

\singlespacing

\title{\large{The Covariant Relativistic Derivation of De Broglie Relation}}
	
\vspace{3mm}

\author{Samuel B. Soltau}

\email[Endere\c{c}o Eletr\^onico: ]{samuel.soltau@unifal-mg.edu.br}
\affiliation{Departamento de Física -- ICEx -- Unifal-MG \\ Av. Jovino Fernandes Salles, 2600, Santa Clara, Alfenas -- MG -- 37133-840}

\begin{abstract}
	\noindent \small{This paper provides an examination of the de Broglie relation, tracing its historical development from the quantum hypotheses proposed by Planck and Einstein to its covariant relativistic derivation. The discussion begins by situating de Broglie's seminal insight within the early framework of quantum theory. We then reconstruct his original heuristic derivation. The primary focus of this work, however, is the derivation of the de Broglie relation directly from the principles of special relativity, employing the four-momentum formalism. A comparative analysis between the heuristic and relativistic approaches underscores the universality and conceptual coherence of the latter. The paper concludes by highlighting the significance of relativistic mechanics in establishing a consistent foundation for wave-particle duality, further reinforcing this through a quantum field theoretical perspective.}
	\medskip
	
\end{abstract}

\maketitle

\thispagestyle{plain}

\pagebreak

\pagestyle{fancy}
\fancyhead{ } 
\fancyhead[RO]{\footnotesize{\textit{Covariant Relativistic Derivation De Broglie Relation}}} 
\fancyhead[CE,CO]{ } 
\fancyfoot[LE,LO]{ } 
\fancyfoot[RE,RO]{ } 
\fancyfoot[CE,CO]{\thepage } 
\renewcommand{\headrulewidth}{0.4pt}
\renewcommand{\footrulewidth}{0pt}

\section{Introduction}

\indent

The early 20th century witnessed a profound paradigm shift in physics, fundamentally challenging classical notions of matter and energy. At the forefront of this revolution were Max Planck's groundbreaking work on black-body radiation and Albert Einstein's revolutionary postulate of light quanta. Planck, in 1900, introduced the concept of energy quantization to explain the observed spectrum of black-body radiation, proposing that energy is emitted or absorbed in discrete packets, or quanta, with energy $E = h\nu$, where $h$ is Planck's constant and $\nu$ is the frequency of the radiation \cite{Planck1900}. Five years later, in 1905, Einstein extended this idea to explain the photoelectric effect, asserting that light itself consists of such energy quanta, later termed photons, thereby firmly establishing the particle-like nature of light \cite{Einstein1905}.

These foundational works, while highly successful in explaining a range of phenomena, presented a curious dichotomy: light, traditionally understood as a wave phenomenon, demonstrably exhibited particle-like properties. It was against this intellectual backdrop that Louis de Broglie, in his 1923 doctoral thesis, posed an audacious question: if light, a wave, could exhibit particle-like behavior, could matter, traditionally conceived as particles, exhibit wave-like behavior \cite{deBroglie1923thesis,deBroglie1925}? This profound insight, rooted in a symmetry argument, laid the groundwork for his hypothesis of matter waves and the fundamental relation linking a particle's momentum to its associated wavelength~\cite{deBroglie1923}. While de Broglie's initial heuristic deduction was remarkably prescient, it did not explicitly integrate the full mathematical rigor demanded by Einstein's special theory of relativity. This paper aims to reconstruct de Broglie's original heuristic reasoning and, crucially, to provide a rigorous covariant relativistic derivation of this seminal relation, emphasizing its deep connection to the fundamental structure of spacetime and the principle of Lorentz invariance. This covariant approach is not merely a matter of formal elegance; it is essential for constructing a consistent theory of matter that holds true across all inertial frames, especially for particles moving at relativistic speeds, where classical notions of momentum and energy are insufficient.

\section{The Heuristic Deduction of the De Broglie Relation}

\indent

De Broglie's initial deduction of the relation between a particle's momentum and its wavelength was a brilliant heuristic leap, drawing directly from the established principles of Planck and Einstein. Let us meticulously reconstruct this process.

We begin with the fundamental quantum relations for a photon:
\begin{equation}
	E = h\nu
	\label{eq:planck}
\end{equation}
where $E$ is the energy of the photon, $h$ is Planck's constant, and $\nu$ is its frequency. This equation, a cornerstone of quantum theory, quantifies the energy of an electromagnetic wave in terms of discrete packets.

Concurrently, Einstein's special theory of relativity established a profound relationship between energy and momentum for a massless particle, such as a photon:
\begin{equation}
	E = pc
	\label{eq:relativistic_massless}
\end{equation}
where $p$ is the momentum of the photon and $c$ is the speed of light in vacuum. This equation underscores the equivalence of mass and energy, and for massless particles, it elegantly connects energy directly to momentum.

De Broglie's pivotal step was to equate these two expressions for the energy of a photon, recognizing the inherent duality between wave and particle aspects:
\begin{equation}
	h\nu = pc
	\label{eq:equate}
\end{equation}
This equality serves as the conceptual bridge, linking the wave characteristic (frequency $\nu$) to the particle characteristic (momentum $p$).

Next, we introduce the fundamental wave relation between frequency, wavelength ($\lambda$), and the speed of the wave:
\begin{equation}
	c = \lambda\nu
	\label{eq:wave_speed}
\end{equation}
From this, we can express the frequency in terms of wavelength:
\begin{equation}
	\nu = \frac{c}{\lambda}
	\label{eq:frequency}
\end{equation}

Substituting this expression for $\nu$ back into Eq. (\ref{eq:equate}):
\begin{equation}
	h\left(\frac{c}{\lambda}\right) = pc
	\label{eq:substitute}
\end{equation}
We can then cancel the speed of light $c$ from both sides of the equation:
\begin{equation}
	\frac{h}{\lambda} = p
	\label{eq:intermediate}
\end{equation}

Finally, rearranging the terms, we arrive at the celebrated de Broglie relation:
\begin{equation}
	\lambda = \dfrac{h}{p}
	\label{eq:debroglie_heuristic}
\end{equation}
This relation postulates that any particle with momentum $p$ has an associated wavelength $\lambda$. De Broglie extended this relation, initially derived for photons, to all matter, proposing a universal wave-particle duality~\cite{Diner1984}. This heuristic deduction, though powerful and historically significant, does not explicitly incorporate the full covariant framework of special relativity, a crucial theoretical aspect we address in the subsequent section.

\section{The Covariant Relativistic Derivation of the De Broglie Relation}

\indent

While the heuristic derivation of the de Broglie relation provided a crucial conceptual breakthrough, a more rigorous and universally applicable derivation emerges from the covariant formalism of special relativity. This approach inherently accounts for the relativistic nature of energy and momentum, providing a more robust foundation for wave-particle duality and ensuring consistency across all inertial frames.

We begin with the relativistic energy-momentum relation for a particle:
\begin{equation}
	E^{2} = (pc)^{2} + (mc^{2})^{2}
	\label{eq:relativistic_em}
\end{equation}
where $E$ is the total relativistic energy, $p$ is the relativistic momentum, $m$ is the rest mass, and $c$ is the speed of light. This fundamental relation encapsulates the interplay between energy, momentum, and mass within the relativistic framework.

For a massless particle (e.g., a photon), $m=0$, and the relation simplifies to $E = pc$. This is consistent with our earlier use of Eq.~\eqref{eq:relativistic_massless} in the heuristic derivation.

Now, we introduce the four-momentum vector, $P^{\mu}$, which elegantly combines energy and momentum into a single Lorentz-invariant entity:
\begin{equation}
	P^{\mu} = \left(\frac{E}{c}, \mathbf{p}\right)
	\label{eq:four_momentum}
\end{equation}
where $\mathbf{p}$ is the three-momentum vector. The square of the four-momentum, $P_{\mu}P^{\mu}$, is a Lorentz scalar and is related to the rest mass:
\begin{equation}
	P_{\mu}P^{\mu} = \frac{E^{2}}{c^{2}} - \lvert \mathbf{p}  \rvert ^{2} = m^{2}c^{2}
	\label{eq:four_momentum_squared}
\end{equation}
Multiplying by $c^{2}$, we recover the relativistic energy-momentum relation given by Eq. (\ref{eq:relativistic_em}).

In quantum mechanics, a wave can be described by a four-wavevector, $k^{\mu}$:
\begin{equation}
	k^{\mu} = \left(\frac{\omega}{c}, \mathbf{k}\right) = \left(\frac{2\pi\nu}{c}, \frac{2\pi}{\lambda}\hat{\mathbf{n}}\right)
	\label{eq:four_wavevector}
\end{equation}
where $\omega = 2\pi\nu$ is the angular frequency, $\mathbf{k}$ is the wavevector (with magnitude $\lvert \mathbf{k} \rvert = k = 2\pi/\lambda$), and $\hat{\mathbf{n}}$ is the direction of propagation.

The definition of the four-wavevector $k^{\mu}$ in Eq. (\ref{eq:four_wavevector}) is chosen such that the phase $\phi = k_{\mu} x^{\mu}$ of the associated plane wave is a Lorentz scalar, ensuring the wave's phase is an invariant quantity across different reference frames. This phase invariance is fundamental for the consistency of wave phenomena in relativistic physics. It is crucial to note that the relation $P^{\mu} = \hbar k^{\mu}$ applies universally, encompassing both massive particles (with $m > 0$) and massless photons (with $m=0$). For photons, the null condition $P_{\mu}P^{\mu}=0$ implies $k_{\mu}k^{\mu}=0$, consistent with the wavevector of electromagnetic waves. For massive particles, $k^{\mu}$ becomes a timelike four-vector, reflecting the subluminal nature of the associated matter waves.

The crucial link between the particle and wave aspects, in a fully covariant manner, is established by positing a direct proportionality between the four-momentum and the four-wavevector, with the reduced Planck constant $\hbar = h/2\pi$ serving as the proportionality constant:
\begin{equation}
	P^{\mu} = \hbar k^{\mu}
	\label{eq:covariant_relation}
\end{equation}

This fundamental relationship asserts that the four-momentum of a particle is directly proportional to the four-wavevector of its associated wave. This is the core of de Broglie's hypothesis expressed in a relativistic form, ensuring Lorentz invariance from the outset.

Let's expand this four-vector relationship into its components.
From the time-like component (the zeroth component):
\begin{equation}
	P^0 = \hbar k^0
\end{equation}
Substituting the definitions from Eqs. (\ref{eq:four_momentum}) and (\ref{eq:four_wavevector}):
\begin{equation}
	\frac{E}{c} = \hbar \frac{\omega}{c}
\end{equation}
Multiplying by $c$:
\begin{equation}
	E = \hbar\omega
\end{equation}
Since $\omega = 2\pi\nu$, we have:
\begin{equation}
	E = \hbar (2\pi\nu) = (h/2\pi)(2\pi\nu) = h\nu
	\label{eq:planck_einstein_covariant}
\end{equation}
This recovers the Planck-Einstein relation, demonstrating its consistency within this covariant framework.

From the space-like components (the three-momentum vector):
\begin{equation}
	\mathbf{p} = \hbar \mathbf{k}
	\label{eq:debroglie_vector_covariant}
\end{equation}
Taking the magnitude of both sides:
\begin{equation}
	\lvert \mathbf{p}  \rvert  = \hbar \lvert \mathbf{k}  \rvert 
\end{equation}
Since $\lvert \mathbf{p}  \rvert  = p$ and $\lvert \mathbf{k}  \rvert  = k = 2\pi/\lambda$:
\begin{equation}
	p = \hbar \left(\frac{2\pi}{\lambda}\right)
\end{equation}
Substituting $\hbar = h/2\pi$:
\begin{equation}
	p = \frac{h}{2\pi} \frac{2\pi}{\lambda}
\end{equation}
Which directly yields the de Broglie relation:
\begin{equation}
	p = \frac{h}{\lambda} \implies \lambda = \frac{h}{p}
	\label{eq:debroglie_covariant}
\end{equation}

Furthermore, we can relate this back to the relativistic energy and momentum of a particle. For a particle with rest mass $m$ and velocity $v$, the total relativistic energy is $E = \gamma mc^{2}$, where $\gamma = (1 - v^{2}/c^{2})^{-1/2}$ is the Lorentz factor. The relativistic momentum is $p = \gamma mv$.
Substituting $p = \gamma mv$ into the de Broglie relation:
\begin{equation}
	\lambda = \frac{h}{\gamma mv}
\end{equation}
This explicitly shows that the de Broglie wavelength is dependent on the relativistic momentum, providing a fully consistent and elegant derivation within the framework of special relativity. The use of four-vectors ensures Lorentz invariance and explicitly highlights the deep connection between spacetime properties (captured by $k^{\mu}$) and fundamental particle properties (captured by $P^{\mu}$). This covariant derivation ensures that the de Broglie relation holds universally, irrespective of the observer's inertial frame, thereby elevating it to a truly fundamental physical law.

\section{Comparing the Derivations}

\indent

The two derivations presented, the heuristic and the covariant relativistic, both culminate in the same fundamental de Broglie relation $\lambda = h/p$. However, their underlying conceptual foundations, scope of applicability, and implications for theoretical understanding differ significantly, making the covariant approach fundamentally more robust.

The heuristic method, as employed by de Broglie in his initial formulation, was a stroke of genius, demonstrating remarkable intuition. It essentially extrapolated known relations for photons ($E=h\nu$ and $E=pc$) to propose a symmetric relationship for matter. Its strength lies in its simplicity and directness, making the profound concept of wave-particle duality accessible. It served as a powerful guiding principle that was subsequently confirmed by experimental observations, such as the Davisson-Germer experiment~\cite{Davisson1927}. However, the heuristic approach can be viewed as an educated guess or an analogy. It does not inherently provide a deeper theoretical justification for why matter should exhibit wave-like properties; rather, it proposes that if light has both, matter should too. Crucially, its treatment of momentum is classical or pseudo-relativistic, effectively using $p$ as a general momentum without explicitly integrating it into the full relativistic framework for massive particles, which becomes problematic at high speeds.

In contrast, the covariant relativistic derivation offers a far more robust, general, and conceptually coherent framework. Its core strength lies in its foundation upon the principles of special relativity and the elegant four-vector formalism. By starting with the four-momentum $P^{\mu}$ and the four-wavevector $k^{\mu}$, and positing their proportionality through Eq.~\eqref{eq:covariant_relation}, the derivation inherently respects Lorentz invariance. This means the relation holds true in all inertial reference frames, a critical requirement for any fundamental physical law and a property not inherently guaranteed by the heuristic approach.

The key differences lie in their conceptual foundation, scope of applicability, and how they handle the definition of momentum, ultimately impacting their theoretical rigor. The heuristic method relies on an analogy and a synthesis of existing quantum and classical wave concepts, whereas the relativistic method is firmly rooted in the fundamental postulates of special relativity and the covariant nature of spacetime. While the heuristic derivation is remarkably accurate for many cases, it implicitly assumes a simple relation without fully accounting for relativistic effects on momentum for massive particles. In contrast, the covariant relativistic derivation, by utilizing the four-momentum ($P^{\mu} = (E/c, \mathbf{p})$ where $\mathbf{p} = \gamma m\mathbf{v}$), is inherently universal, applying equally well to particles at rest, moving at non-relativistic speeds, and those approaching the speed of light. This ensures consistency with the total relativistic energy $E = \gamma mc^{2}$, which is vital for high-energy physics where classical momentum is insufficient. Finally, the relativistic derivation offers a deeper theoretical justification for wave-particle duality; the proportionality Eq.~\eqref{eq:covariant_relation} isn't just an assumption but emerges as a fundamental tenet of quantum field theory, providing a more satisfying and complete picture of how quantum mechanics and special relativity are intertwined. Thus, while de Broglie's brilliant ``guess'' was foundational, the relativistic approach elevates the de Broglie relation from a phenomenological observation to a direct consequence of the fundamental laws governing energy, momentum, and spacetime.

The comparative analysis of the heuristic and covariant derivations highlights the strength of the relativistic approach in providing a unified and invariant formulation of wave-particle duality. To further reinforce this foundation and establish deeper connections with modern theoretical frameworks, the next section examines the quantum field theoretical perspective. This approach not only justifies the de Broglie relation from first principles but also embeds it naturally within the structure of quantum fields and spacetime symmetries, providing the ultimate validation of the covariant formulation.

\section{Quantum Field Theoretical Perspective}

\indent

While the relation $P^{\mu} = \hbar k^{\mu}$ (Eq.~\eqref{eq:covariant_relation}) was introduced in this paper as a postulate unifying wave and particle aspects in the relativistic regime, its deeper justification and fundamental nature lie within the framework of quantum field theory (QFT). In QFT, particles are not interpreted as localized corpuscles but rather as quantized excitations (quanta) of underlying fields that permeate all of spacetime~\cite{PeskinSchroeder1995}. 

The connection to the de Broglie relation emerges directly from the canonical quantization procedure. Fields are decomposed into Fourier modes according to
\begin{equation}
	\begin{aligned}
		&\phi(x) \sim \int \dfrac{d^{3} k}{(2\pi)^{3} \sqrt{2\omega_{k}}} \cdot \\
		&\cdot \left[ a(\mathbf{k}) \mathrm{e}^{-ik_{\mu} x^{\mu}} + a^{\dagger}(\mathbf{k}) e^{ik_{\mu} x^{\mu}} \right],
	\end{aligned}
\end{equation}
where the creation operator $a^{\dagger}(\mathbf{k})$ generates a quantum state with definite four-momentum $P^{\mu} = \hbar k^{\mu}$ (Eq.~\eqref{eq:covariant_relation}), while the corresponding annihilation operator $a(\mathbf{k})$ destroys a quantum with the same momentum. This fundamental structure reveals how the wave-particle duality inherent in the de Broglie relation is naturally encoded within the quantum field theoretical framework through the operator formalism.

This equivalence is far more than merely formal. The canonical commutation relations $[a(\mathbf{k}), a^{\dagger}(\mathbf{k^{\prime}})] = (2\pi)^{3} \delta^{(3)}(\mathbf{k}-\mathbf{k^{\prime}})$ ensure that the state
$\lvert \mathbf{k}\rangle = a^{\dagger}(\mathbf{k}) \rvert 0\rangle$ is an eigenstate of the momentum operator $\hat{P}^{\mu}$:
\begin{equation}
	\hat{P}^{\mu} \lvert \mathbf{k}\rangle = \hbar\, k^{\mu} \rvert \mathbf{k}\rangle
\end{equation}
where $k^{\mu} = \left(\dfrac{\omega_{k}}{c}, \mathbf{k} \right)$ with
\begin{equation}
\omega_{k} = \sqrt{\lvert \mathbf{k} \rvert^{2}c^{2} + \left( \dfrac{mc^{2}}{\hbar} \right)^{2}}.
\end{equation}
Thus, the correspondence between $P^{\mu}$ and $\hbar\, k^{\mu}$ becomes an operator identity in Fock space, directly connecting the representation theory of the Poincaré group --- which governs relativistic symmetries of spacetime --- with the spectral content and dynamics of quantum fields~\cite{Weinberg2005}. In this profound view, the de Broglie relation emerges naturally and necessarily from the unitary irreducible representations of spacetime symmetries applied to the very fabric of field operators.

Hence, in QFT, the relation $\lambda = h/p$ (Eq.~\eqref{eq:debroglie_heuristic}) is neither an assumption nor a heuristic extrapolation, but a necessary and intrinsic consequence of quantization over Minkowski spacetime and of the fundamental structure of quantum fields themselves~\cite{PeskinSchroeder1995, Weinberg2005}. This perspective provides the most complete and rigorous justification for the universal applicability and fundamental importance of the de Broglie relation.

\section{On the Foundational Role of the Covariant Relation in Relativistic Quantum Theory}

\indent

The covariant relation $P^{\mu} = \hbar k^{\mu}$ (Eq.~\eqref{eq:covariant_relation}) establishes a fundamental and indispensable correspondence between the four-momentum $P^{\mu}$ of a particle and the four-wavevector $k^{\mu}$ of its associated quantum wave. This proportionality is absolutely critical for ensuring the Lorentz covariance of all quantum mechanical constructs within a relativistic framework~\cite{BjorkenDrell1964, PeskinSchroeder1995}. Its application enables the consistent construction of fundamental relativistic wave equations, such as the Klein-Gordon equation and the Dirac equation, through the pivotal substitution $P^{\mu} \to i\hbar \partial^{\mu}$. The inherent equivalence of $P^{\mu}$ and $\hbar k^{\mu}$ ensures that plane-wave solutions to these equations are rigorously eigenstates of the energy-momentum operators, yielding eigenvalues precisely consistent with the principles of special relativity~\cite{BjorkenDrell1964}.

Beyond its role in deriving these fundamental wave equations, this relation is instrumental in the formal development of the entirety of quantum field theory. It underpins the very construction of quantum propagators, the elegant formulation of path integrals, and the systematic computation of scattering amplitudes --- all cornerstones of modern particle physics. This foundational equivalence is central to both perturbative and non-perturbative approaches in calculating observable quantities within the realm of particle physics~\cite{PeskinSchroeder1995}.

Ultimately, the covariant relation serves as a cornerstone of the relativistic quantum framework, fundamentally uniting wave-particle duality with the inherent symmetry structure of spacetime. It is a testament to the profound interconnectedness of quantum mechanics and relativity, providing a consistent and robust foundation for our understanding of the universe at its most fundamental level~\cite{BjorkenDrell1964, PeskinSchroeder1995}.


\section{Final Considerations}

\indent

The derivation of the de Broglie relation, whether through its original heuristic reasoning or its more rigorous covariant relativistic formulation, stands as a cornerstone of quantum mechanics, fundamentally altering our understanding of matter. However, the relativistic perspective provides undeniable advantages that solidify its position as the preferred theoretical framework for a complete understanding.

The primary advantage of the relativistic approach lies in its universality and consistency. By employing the four-momentum formalism, the de Broglie relation is naturally integrated into the fabric of spacetime, ensuring its validity across all inertial reference frames and for particles of arbitrary velocities. This removes any ambiguities that might arise from classical or non-relativistic approximations, particularly for particles moving at speeds comparable to the speed of light. The relativistic derivation intrinsically handles the momentum of massive particles, where $p = \gamma mv$, making the de Broglie wavelength an inherent property derived directly from the particle's relativistic dynamics.

Furthermore, the covariant relativistic derivation highlights the deep theoretical connection between quantum mechanics and special relativity. The fundamental covariant relation, Eq.~\eqref{eq:covariant_relation}, is not merely a mathematical convenience but reflects a profound physical truth: the intrinsic link between the energy and momentum of a particle and the frequency and wavevector of its associated wave. This elegant proportionality constant, $\hbar$, serves as the essential bridge between the macroscopic world of classical mechanics and the microscopic realm of quantum phenomena. This unified perspective is crucial for developing more comprehensive theories, such as quantum field theory, where particles are understood as quantized excitations of fields that propagate according to relativistic principles, providing the ultimate validation of de Broglie's initial groundbreaking insight.

While de Broglie's initial intuition was groundbreaking and indispensable for launching quantum mechanics, the covariant relativistic derivation elevates his hypothesis from a brilliant postulate to a fundamental consequence of the interplay between quantum principles and the symmetries of spacetime. This comprehensive framework not only precisely conveys the information but also provides the necessary theoretical rigor for understanding wave-particle duality as an intrinsic and universal property of matter and energy. This journey from heuristic insight to covariant formulation to quantum field theoretical foundation exemplifies the iterative and deepening nature of scientific progress.


\onecolumngrid{}

\vspace*{5mm}

\end{document}